\documentclass[3p,review]{elsarticle}
\usepackage{lineno,rotating}
\modulolinenumbers[5]
\journal{Nuclear Physics A}
\begin{document}
\begin{frontmatter}
\title{Extended systematics of alpha decay half lives for exotic superheavy nuclei}
\author[mymainaddress]{A. I. Budaca\corref{mycorrespondingauthor}}
\cortext[mycorrespondingauthor]{Corresponding author}
\ead{abudaca@theory.nipne.ro}
\author[mymainaddress]{R. Budaca}
\author[mymainaddress]{I. Silisteanu}
\address[mymainaddress]{Department of Theoretical Physics, Horia Hulubei National Institute for
Physics and Nuclear Engineering, Reactorului 30, RO-077125,
POB-MG6, Bucharest Magurele, Romania}

\begin{abstract}
The experimentally available data on the $\alpha$ decay half lives and $Q_{\alpha}$ values for 96 superheavy nuclei are used to fix the parameters for a modified version of the Brown empirical formula through two fitting procedures which enables its comparison with similar fits using Viola-Seaborg and Royer formulas. The new expressions provide very good agreement with experimental data having fewer or the same number of parameters. All formulas with the obtained parameters are then extrapolated to generate half lives predictions for 125 unknown superheavy $\alpha$ emitters. The nuclei where the employed empirical formulas maximally or minimally diverge are pointed out and a selection of 36 nuclei with exceptional superposition of predictions was made for experimental reference.
\end{abstract}
\end{frontmatter}

\section{Introduction}

The superheavy nuclei (SHN) pose a continuous challenge for experimental studies due to their elusive synthesis and still widely unknown properties. This is where the theoretical studies come to help fueled also by the prospect of understanding such nuclear properties as shell stability, magic numbers and island of stability. Both theoretical and experimental facets of SHN are extensively presented in Ref.\cite{HoffRMP}. As $\alpha$ emission is the predominant decay mode of unstable nuclei, it obviously represents the best tool to study SHN. Moreover, it provides a very efficient way to identify new elements through the observation of the $\alpha$ decay chain from an unknown nucleus up to its end nucleus, which disintegrates by spontaneous fission. Basically, the experimental output consists of $\alpha$ decay kinetic energy and half-lives. The prediction of these quantities provides a major support for experimental design, consequently it is also the main goal of the theory.

Based on the simplified picture of the $\alpha$ decay as a semiclassical one-dimensional process, simple but rather striking correlations between the $\alpha$ decay $Q_{\alpha}$ values and half lives emerge which provide a similar good agreement with experiment like the more sophisticated models. As the amount of the decay data continuously grows, simple and fast calculations are necessarily required. In this context, the empirical formulas which usually relates $\log{T_{\alpha}}$ with the $Q_{\alpha}$ values and the $(Z,A)$ of the parent nucleus offer easily obtainable updated results. The first correlation of this type was registered by Geiger and Nuttall \cite{GN}, who found that $\log{T_{\alpha}}$ depends linearly on $Q_{\alpha}^{-1/2}$. Various generalizations \cite{TN,Viola,Soby,Brown,Moller,Royer,Horoi,Park,Poenaru,Medeiros,SobiPark,Ni,Santosh,Qi1,Qi2,Denisov,RenPRC} of the Geiger-Nuttall law systematically compared in Ref.\cite{Wang} share this $Q_{\alpha}$ dependence. This relation is also extrapolated to heavy cluster \cite{Horoi,Ni,Po1,Po2,Bala,Ren} and proton \cite{Delion,Medeirosp,Dongp} decays. This is not surprising, because it originates from the common quantum penetration hypothesis. The coefficients of this linear relation are however not universal, varying between different decay chains. The first successful generalization of the Geiger-Nuttall law, which transcends the decay chains, came with the work of Viola and Seaborg \cite{Viola} who, inspired by the work of Gallagher and Rasmussen \cite{Gala}, considered for the slope and intercept parameters a linear dependence on the charge number of the daughter nucleus. Later on, Sobiczewski and collaborators \cite{Soby} extrapolated the formula for deformed SHN providing a set of parameters which remains actual even today. As a consequence, even though the formula was proposed more than 50 years ago, it represents in its updated form \cite{Soby} one of the reference empirical correlation for the $\alpha$ decay due to its simple form and unmatched predictive power, taking into account the number of parameters involved. The only downfall of the formula is the lack of physical justification for such a dependence of the coefficients. However, taking in consideration higher order terms in the expansion of the Wentzel-Kramers-Brillouin (WKB) barrier penetration factor, it was observed that the plots versus $Z_{d}Q_{\alpha}^{-1/2}$ are better suited for data description, supporting thus the Viola-Seaborg correlation formula.

Going further, Brown \cite{Brown} considered an interpolation between the Geiger-Nuttall law and the $Z_{d}Q_{\alpha}^{-1/2}$ dependence, obtaining that the best representation of data is achieved by the linear dependence of the $\log{T_{\alpha}}$ on $Z_{d}^{0.6}Q_{\alpha}^{-1/2}$ quantity. Including similar $(Z,N)$ parity varying hindrance terms as in \cite{Viola} proved the Brown formula to be an extremely powerful tool for describing the $\alpha$ decay properties of SHN \cite{Budaca1,Budaca2,Budaca3,Budaca4,Budaca5,Budaca6,Budaca7}. Moreover, it was also generalized for cluster decays \cite{Horoi}, by additionally fitting the power of $Z_{d}$ which surprisingly remained essentially the same.

Here one will show that for SHN a different power of $Z_{d}$ must be considered in order to have the best agreement with experiment. To do this, we gather the available experimental data regarding the $Q_{\alpha}$ and $\log{T_{\alpha}}$ values of SHN and fit it against a modified version of the Brown formula through two distinct schemes. Such that the results obtained with both procedures are compared with experimental data available for 96 SHN as well as similar fits in respect to other two most successful correlation functions, Viola-Seaborg \cite{Viola} and Royer \cite{Royer}. The obtained parameters are further used to generate predictions for 125 unknown $\alpha$ emitters, whose $Q_{\alpha}$ values are determined through a simple five-parameter formula deduced from the Liquid Drop Model (LDM) considerations \cite{DongPRC,DongPRL}. As this method of $Q_{\alpha}$ calculus produce very good results when extrapolated to immediate neighboring nuclei \cite{DongPRL}, the nuclei considered for predictions are adjacent to those with experimentally available data. Regions of agreement with experiment as well as specific differences between predictions are identified through various graphical representations, bringing us to the main scope of the paper which is to offer reliable half-live predictions for unknown SHN.

\renewcommand{\theequation}{2.\arabic{equation}}
\setcounter{equation}{0}
\section{Empirical $\alpha$-decay half lives estimates}

In what follows one will shortly present the two most successful empirical formulas as well as a modified version of the Brown formula. For avoiding inconsistencies, when the atomic, neutron and mass numbers are not accompanied by an index, they are referring to the parent nucleus.

\subsection{Viola-Seaborg formula}
The well known Viola-Seaborg (VS) formula \cite{Viola} writes as
\begin{equation}
\log{T_{\alpha}^{VS}}=(aZ+b)Q_{\alpha}^{-1/2}+(cZ+d)+h_{Z-N}^{VS},
\label{vi-se}
\end{equation}
where $Q_{\alpha}$ is the decay energy in MeV units, $Z$ is the charge number of the parent nucleus, $h_{Z-N}^{VS}$ is an even-odd hindrance term, while $a,b,c$ and $d$ are fitting parameters. The commonly used values for the parameters are those from \cite{Soby}: $a=1.66175$, $b=-8.5166$, $c=-0.20228$, $d=-33.9069$ and
\begin{equation}
h_{Z-N}^{VS}=\left\{\begin{array}{ll}
    0,\,\,\,\,\,\,\,\,\,\,\,\,\,(e-e),\\
    1.066,\,\,(e-o),\\
    0.772,\,\,(o-e),\\
    1.114,\,\,(o-o),\\
\end{array}\right.
\label{hsoby}
\end{equation}
are values from the original paper of Viola and Seaborg \cite{Viola}. $(e-e)$ means $Z$ and $N$ even, $(e-o)$ - $Z$ even and $N$ odd, $(o-e)$ - $Z$ odd and $N$ even, $(e-e)$ - $Z$ and $N$ odd. Other sets of parameters are constantly provided by fits on updated and new experimental data \cite{DongRen} or different sets of highly precise data \cite{Dasgupta}.

\subsection{Royer formula}
Another successful empirical formula was proposed by Royer in Ref.\cite{Royer}. Besides the dependence of $\log{T_{\alpha}}$ on $Z$ and $Q_{\alpha}$ it also incorporates a variance with the atomic mass number $A$:
\begin{equation}
\log{T_{\alpha}^{R}}=a Z Q_{\alpha}^{-1/2}+b A^{1/6}Z^{1/2}+c.
\end{equation}
Although the Royer (R) formula can be used for global fits regardless of the $(Z,N)$ parity of the parent nucleus, it is usually referred to as a 12-parameter correlation, with distinct coefficients $a,b$ and $c$ for each $(Z,N)$ parity combination. The original set of parameters \cite{Royer}
\begin{equation}
(a,b,c)_{Z-N}^{R}=\left\{\begin{array}{ll}
    1.5864,-1.1629,-25.31,\,\,(e-e),\\
    1.5848,-1.0859,-26.65,\,\,(e-o),\\
    1.5920,-1.1423,-25.68,\,\,(o-e),\\
    1.6971,-1.1130,-29.48,\,\,(o-o),
\end{array}\right.
\label{Royset}
\end{equation}
was established by fitting experimental data for 373 cumulated nuclei. Other optimised fits with this formula showed its superior experimental data reproduction \cite{Dasgupta}, making it along with VS formula one of the best empirical relation between the properties of the $\alpha$ decay. It is worth mentioning that this empirical correlation is actually a special case of the unified formula of half-lives for $\alpha$ decay and cluster radioactivity proposed in Ref.\cite{Ni}, and which was shown in Ref.\cite{Wang} that leads to the smallest rms values when fitting all known $\alpha$ decay data.

\subsection{Modified Brown formula}
The majority of experimental $\log{T_{\alpha}}$ values when expressed as function of the $Z_{d}^{0.6}{Q_{\alpha}^{-1/2}}$ quantity were observed in Ref.\cite{Brown} to fall on a single straight line defined as:
\begin{equation}
\log{T_{\alpha}^{B}}=9.54(Z-2)^{0.6}Q_{\alpha}^{-1/2}-51.37,
\end{equation}
where $Z$ is the charge number of the parent nucleus. The parameters where fixed by fitting the available experimental data at that time. Recently, the formula was extrapolated to fits for SHN in Refs. \cite{Budaca1,Budaca2,Budaca3,Budaca4,Budaca5,Budaca6,Budaca7} where the authors also added an even-odd hindrance term and obtained updated parameters for this region of the nuclide chart. Although the $0.6$ power is numerically supported by the semiclassical WKB calculation, it was determined from the fits to experimental data \cite{Brown,Horoi}. Due to the great flux of new and improved experimental data especially in what concerns SHN, this value must be revised. For this, one will adopt in this work the modified Brown (mB) formula
\begin{equation}
\log{T_{\alpha}^{mB}}=a(Z-2)^{b}Q_{\alpha}^{-1/2}+c,
\label{brown}
\end{equation}
where $a,b,c$ are fitting parameters. It is worth to mention here that, contrary to the VS formula where the change from $Z$ to $Z-2$ just redefines the involved parameters, here it leads to a completely different function. The same change is valid also in case of the R formula.

\renewcommand{\theequation}{3.\arabic{equation}}
\setcounter{equation}{0}
\section{Determination of $Q_{\alpha}$ values}

The $Q_{\alpha}$ values for measured superheavy $\alpha$ emitters are extracted from the $\alpha$ particle kinetic energy $E_{\alpha}$. These two quantities are related through the formula:
\begin{equation}
Q_{\alpha}=\frac{A}{A-4}E_{\alpha}+\left[6.53(Z-2)^{7/5}-8.0(Z-2)^{2/5}\right]\cdot10^{-5}\,\,\textrm{MeV},
\label{QE}
\end{equation}
amounting the standard recoil and electron shielding corrections \cite{Rasmusen,Pearl} and where $A$ and $Z$ are the mass and atomic numbers of the parent nucleus.

In what concerns $Q_{\alpha}$ predictions, these are calculated with the aid of a simple but very efficient five-parameter formula derived from LDM considerations \cite{DongPRC,DongPRL}:
\begin{eqnarray}
Q_{\alpha}^{Th}(\textrm{MeV})&=&\alpha ZA^{-4/3}(3A-Z)+\beta\left(\frac{N-Z}{A}\right)^{2}\nonumber\\
&&+\gamma\left[\frac{|N-152|}{N}-\frac{|N-154|}{N-2}\right]\nonumber\\
&&+\delta\left[\frac{|Z-110|}{Z}-\frac{|Z-112|}{Z-2}\right]+\epsilon.
\label{MDong}
\end{eqnarray}
The first two terms are the contributions coming from the LDM Coulomb energy and symmetry energy, respectively, while the next two account for the neutron and proton shell effects of $N=152$ and respectively $Z=110$.  The parameters involved were determined in \cite{DongPRC} by fitting 154 experimental $Q_{\alpha}$ data points and have the values $\alpha=0.9373$, $\beta=-99.3027$, $\gamma=16.0363$, $\delta=-21.5983$ and $\epsilon=-27.4530$. The resulted deviations recommend the same values for reliable predictions corresponding to distant SHN. The choice of this formula for predictions is also supported by the calculations made in Ref. \cite{Wang}, where was shown that the above formula perform better than most recent fully LDM \cite{RoyerQ}, macroscopic-microscopic \cite{Liu,Bhagwat}, fully microscopic \cite{QiQ}, mean field \cite{Goriely,Zheng,Qu} or infinite nuclear matter model \cite{Nayak} based methods, being outmatched only by another LDM based fitting expression \cite{DongQ} and a macroscopic-microscopic model estimation \cite{WangQ}, which however uses eight and respectively nine parameters.

\renewcommand{\theequation}{4.\arabic{equation}}
\setcounter{equation}{0}
\section{Numerical applications}

\subsection{Experimental data fits}

In order to have a truthful comparison with the aforementioned VS and R formulas one will consider two fitting schemes for the modified Brown formula. In accordance to the structure of the VS formula, the first modified Brown fit (mB1) will have the parameters $a$, $b$ and $c$ parity independent with an additional hindrance term differentiated by parity:
\begin{equation}
\log{T_{\alpha}^{mB1}}=a(Z-2)^{b}Q_{\alpha}^{-1/2}+c+h_{Z-N}^{mB1},
\label{mB1}
\end{equation}
totaling thus six fitting parameters in comparison to the seven ones used in VS formula.

Following the same reasoning, in case of the second fitting scheme for the modified Brown formula (mB2), all three original parameters $a$, $b$ and $c$ have four distinct sets of values corresponding to each $(Z,N)$ parity. Therefore the corresponding fitting function reads as:
\begin{equation}
\log{T_{\alpha}^{mB2}}=a_{Z-N}(Z-2)^{b_{Z-N}}Q_{\alpha}^{-1/2}+c_{Z-N},
\label{mB2}
\end{equation}
and it has the same number of fitting parameters (12) as the R formula.

The parameters of the empirical formulas are usually established by fitting large amount of data which however is subjected to significant errors and spread over regions of nuclide chart with different properties. The predictions of resulting formulas lose from their reliability in this way. This means that the relation between half live time, reaction energy and number of constituent nucleons is more complex than that. Indeed, the agreement with experiment is substantially improved if these nuclear characteristics are correlated with additional experimental observables. The use of experimental charge radii in a generalized density-dependent cluster model for long lived $\alpha$ emitters \cite{RenPLB} is one example. Fits on most precise experimental data \cite{Dasgupta} represent an alternative approach which in turn lack the same statistics. It is well known that the existence of SHN nuclei is owed to the microscopic shell structure. On the other hand they are also vastly influenced by collective degrees of freedom which lead to a greater variety of shapes than in the lighter $\alpha$ emitters, and emergence of such rare phenomena as meta-stable states, shape coexistence and isomers \cite{Natur}. Therefore, in order to have empirical correlations with higher power of predictability it is necessary to fix the involved parameters in specific regions. Such a region is defined by the concourse of all existing SHN. Following this logic, the experimental data on $E_{\alpha}$ and $\log{T_{\alpha}}$ values for 96 SHN gathered from Refs.\cite{Morita,Og1,Og2,Dvorak,Og3,Elliot,Dullman,Hofmann,Og4,Og5} are fitted with VS, R, mB1 and mB2 formulas. The experimental $Q_{\alpha}$ values are extracted from the measured kinetic energy of the $\alpha$ particle $E_{\alpha}$ by means of (\ref{QE}). The fits are judged by the root mean square (rms):
\begin{equation}
\textrm{rms}=\sqrt{\frac{1}{N_{nucl}}\sum_{i=1}^{N_{nucl}}\left(\log{\frac{T_{Th}^{i}}{T_{Exp}^{i}}}\right)^{2}},
\label{rms}
\end{equation}
where $N_{nucl}$ represents the number of nuclei considered in the fitting procedure. Other quantities of interest are the mean deviation
\begin{equation}
\delta=\frac{1}{N_{nucl}}\sum_{i=1}^{N_{nucl}}\left|\log{\frac{T_{Th}^{i}}{T_{Exp}^{i}}}\right|,
\end{equation}
and the index $F=10^{\delta}$ \cite{Park,SobiPark}.

The results of the fits, regarding the obtained parameters and associated statistical quantities are given in Table \ref{par}. One excluded from the above fits three experimental data points corresponding to $^{263}108$, $^{266}109$ and $^{281}110$ which are well known from the literature that have big fall outs from various theoretical estimations. As it is shown in Table \ref{fall} this is also the case of the present empirical calculations where all four theoretical half live estimations differ from the corresponding experimental values by few orders of magnitude. These large discrepancies are usually ascribed to experimental errors. As a matter of fact, in \cite{Budaca4} were presented big discrepancies between theory and experiment also for $^{282}$Rg and $^{290}115$ and their agreement with experiment was shown, in \cite{Budaca7}, to be improved after a new series of experiments.

\setlength{\tabcolsep}{6.1pt}
\begin{sidewaystable}[th!]
\caption{The new parameters of the VS and mB1 formulas resulting from the first fitting procedure and the new $(Z,N)$ parity differentiated parameters corresponding to R and mB2 fits. The corresponding statistical quantities rms, $\delta$ and $F$ are also listed. In both VS and mB1 cases $h_{e-e}=0$.}
\label{par}
\begin{center}
\begin{tabular}{ccrrrcccccccc}
\hline\hline\noalign{\smallskip}
Formula&$(Z-N)$&$a$&$b$&$c$&$d$&$h_{e-o}$&$h_{o-e}$&$h_{o-o}$&rms&$\delta$&$F$&Nr. of SHN\\
\noalign{\smallskip}\hline\noalign{\smallskip}
VS&all&1.5744 &-23.3922&-0.2746 &-18.2236&0.4479&0.5780&0.7944&0.3951&0.3189&2.0840&96\\
mB1&all&13.0705&0.5182  &-47.8867&        &0.4666&0.6001&0.8200&0.3972&0.3192&2.0855&96\\
\noalign{\smallskip}\hline\noalign{\smallskip}
R &$(e-e)$& 1.6672&-1.2216&-26.3843&&&&&0.0697&0.0186&1.0438&13\\
mB2&$(e-e)$&10.8238& 0.5966&-56.9785&&&&&0.0646&0.0175&1.0411&13\\
R &$(e-o)$& 1.4763&-1.3523&-15.8306&&&&&0.2311&0.1071&1.2796&34\\
mB2&$(e-o)$&14.7747& 0.5021&-49.7080&&&&&0.2313&0.1033&1.2686&34\\
R &$(o-e)$& 1.1499&-1.0402&-12.6186&&&&&0.1281&0.0489&1.1192&18\\
mB2&$(o-e)$&11.1462& 0.5110&-39.0096&&&&&0.1318&0.0504&1.1230&18\\
R &$(o-o)$& 1.2451&-1.2134&-11.1310&&&&&0.2314&0.1092&1.2860&31\\
mB2&$(o-o)$&14.7405& 0.4666&-41.7227&&&&&0.2274&0.1082&1.2828&31\\
\noalign{\smallskip}\hline\hline
\end{tabular}
\end{center}
\end{sidewaystable}

Before making a comparison between fits results, it is instructive to register the differences with the corresponding original sets of parameters. Thus, comparing to the values reported in Ref.\cite{Soby}, the actual values for the VS formula are similar in case of $a$ and $c$ parameters, while $b$ and $d$ differ directly and inversely by factors 2.75 and 1.86, respectively. In what concerns the hindrance term $h_{Z-N}^{VS}$, the present fits offer values which are overall lower in respect to those from (\ref{hsoby}). Moreover the ordering of these factors by magnitude is $(e-o)<(o-e)<(o-o)$ instead of $(o-e)<(e-o)<(o-o)$ as prescribed by (\ref{hsoby}) and Ref.\cite{Viola}. There are two reasons for this change: first is the fact that the fits are performed only on SHN and the second is the different number of experimental data points for $(e-o)$ and $(o-e)$ nuclei (see table \ref{par}). The differences are more drastic in what concerns R formula. Indeed, only the $(e-e)$ parameters resemble the original values \cite{Royer}, for the rest of parities the most affected is the $c$ parameter whose absolute value is smaller by an average factor of two. The parameters $a$ and $c$ obtained here in all cases for the modified Brown formula are in agreement with those reported in Refs. \cite{Budaca1,Budaca2,Budaca3,Budaca4,Budaca5,Budaca6,Budaca7}, while the power $b$ is reduced with respect to the original 0.6 power of the Brown formula \cite{Brown}. As a mater of fact, excepting the value corresponding to mB2 $(e-e)$ which is very close to the 0.6 value, the rest of the fitted powers are more close to the 0.5 value, with the mB2 $(o-o)$ even lower than that. Note also that the inverse ordering of the $h_{Z-N}$ factors obtained in VS fits is maintained.  Another interesting aspect is that all hindrance parameters of the Bm1 formula are shifted by $\approx$0.02 in regard to the values obtained with the VS fit.

\setlength{\tabcolsep}{8.5pt}
\begin{table*}[th!]
\caption{Experimentally measured $\alpha$ decay half lives for SHN known to disagree with various theoretical estimations are compared with the results provided by VS, R and the two versions of mB empirical formulas with the parameters listed in Table \ref{par}.}
\label{fall}
\begin{center}
\begin{tabular}{cccrrrrrrr}
\hline\hline\noalign{\smallskip}
 $Z$& $N$& $A$& $E_{\alpha}$~~~& $Q_{\alpha}$~~~& $\log{T_{\alpha}^{exp}}$& $\log{T_{\alpha}^{VS}}$& $\log{T_{\alpha}^{mB1}}$& $\log{T_{\alpha}^{R}}$& $\log{T_{\alpha}^{mB2}}$\\
    &    &    &   (MeV)       &     (MeV)     & (s)~~~~                 & (s)~~~~                &  (s)~~~~                &  (s)~~~~              &  (s)~~~~                \\
\noalign{\smallskip}\hline\noalign{\smallskip}
 108 & 155 & 263 & 10.90 & 11.11 & -0.131 & -3.441 & -3.473 & -3.573 & -3.624 \\
 109 & 157 & 266 & 11.74 & 11.96 & -2.770 & -4.508 & -4.506 & -4.021 & -4.006 \\
 110 & 171 & 281 &  8.73 &  8.90 &  1.104 &  2.227 &  2.161 &  2.302 &  2.268 \\
  \noalign{\smallskip}\hline\hline
\end{tabular}
\end{center}
\end{table*}

\begin{figure}[ht!]
 \centering
 \includegraphics[clip,trim = 23mm 10mm 30mm 20mm,width=0.65\textwidth]{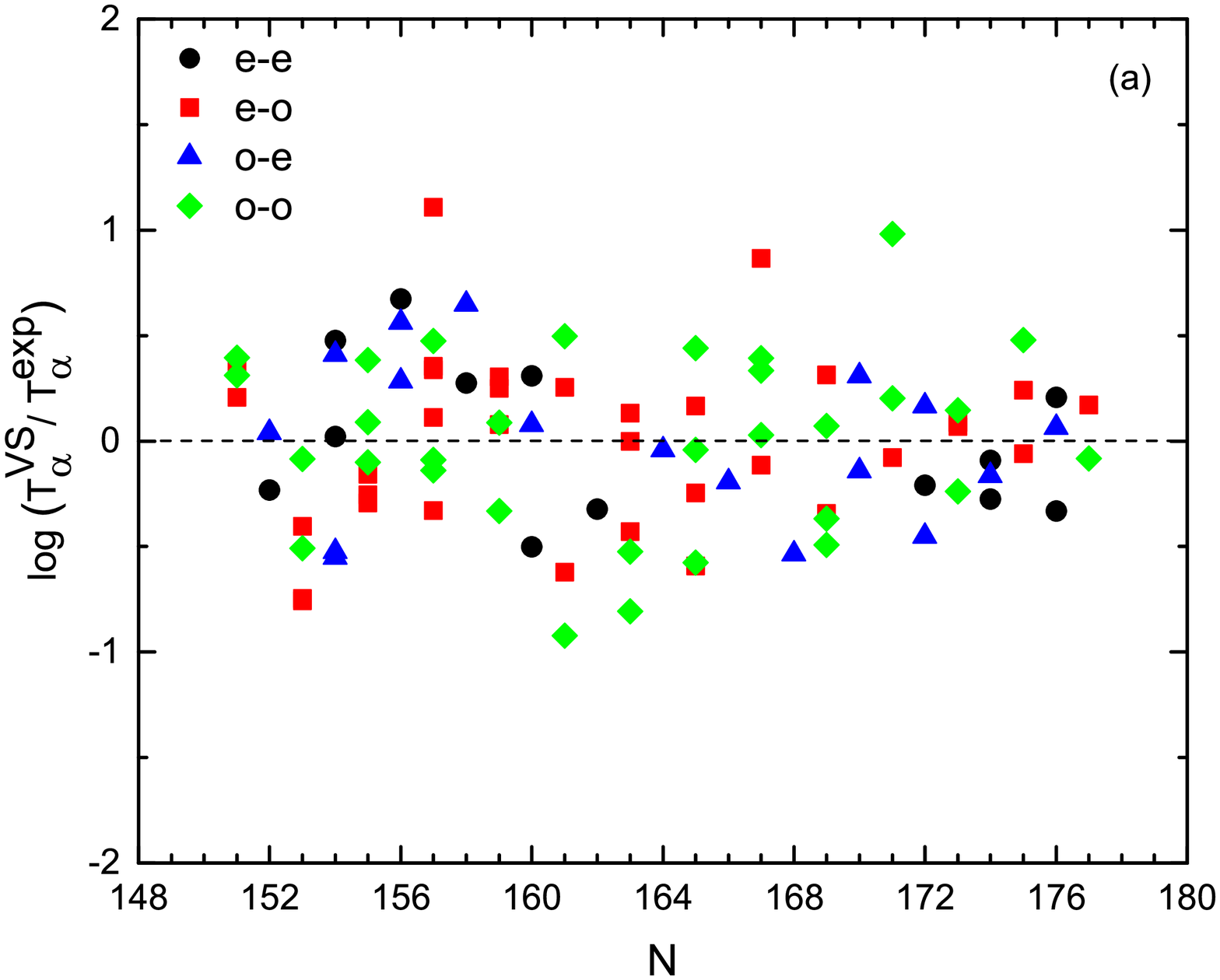}\\
 \includegraphics[clip,trim = 23mm 10mm 30mm 20mm,width=0.65\textwidth]{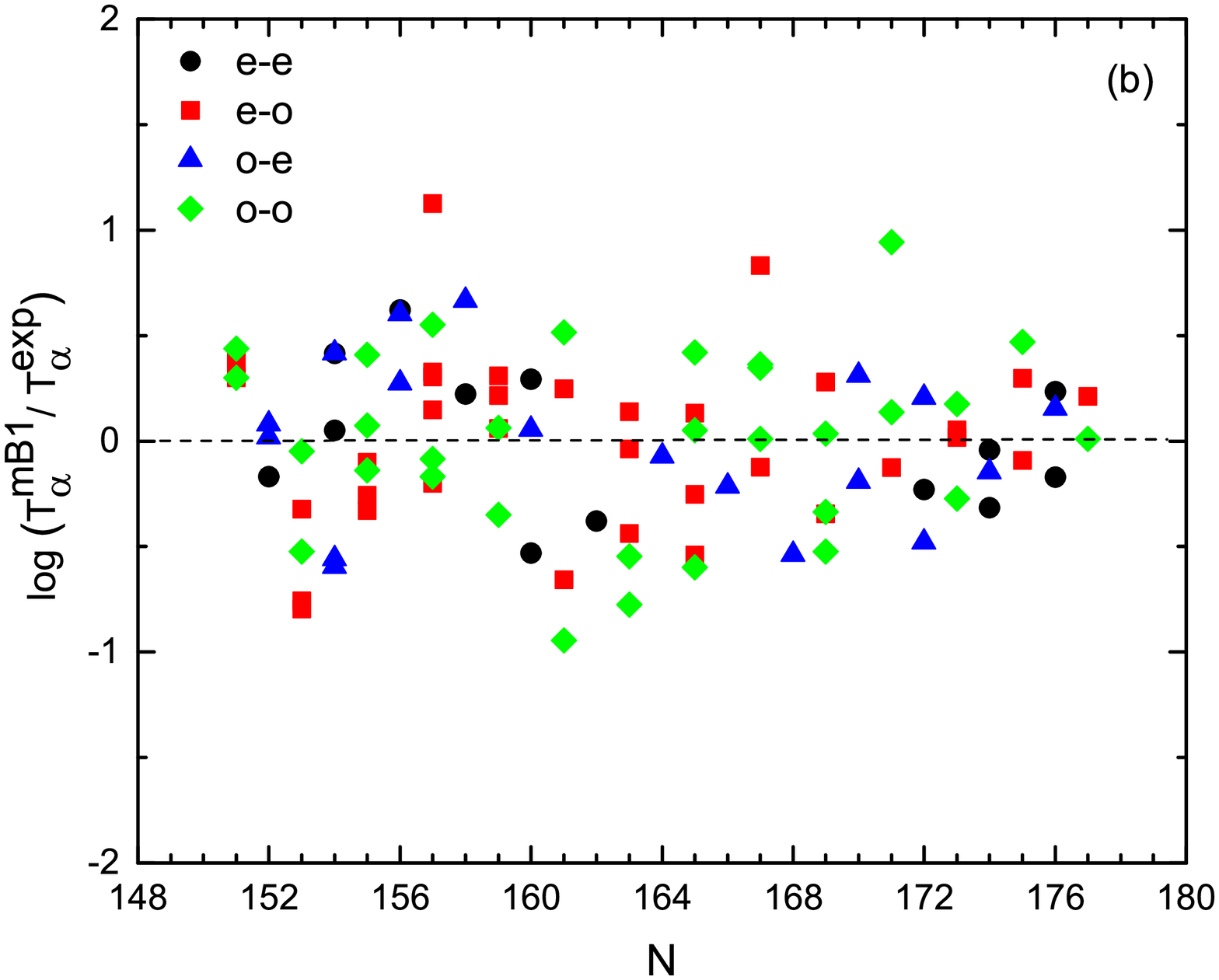}
 \caption{The deviations of the VS (a) and mB1 (b) formula predictions from experimental data as function of neutron number $N$ for 96 data points.}
 \label{VSEXPB1}
\end{figure}

\begin{figure}[ht!]
 \centering
 \includegraphics[clip,trim = 23mm 10mm 30mm 20mm,width=0.65\textwidth]{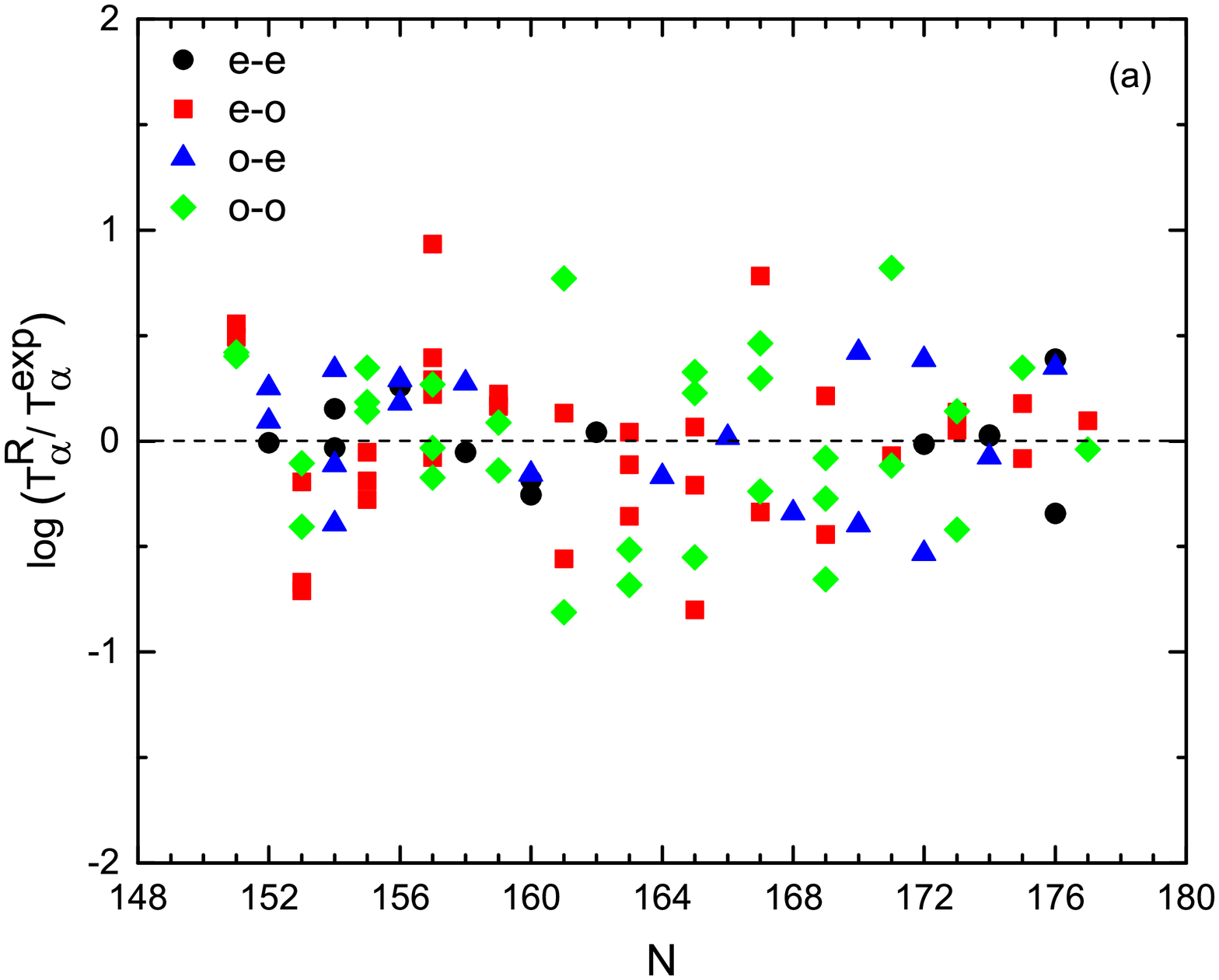}
 \includegraphics[clip,trim = 23mm 10mm 30mm 20mm,width=0.65\textwidth]{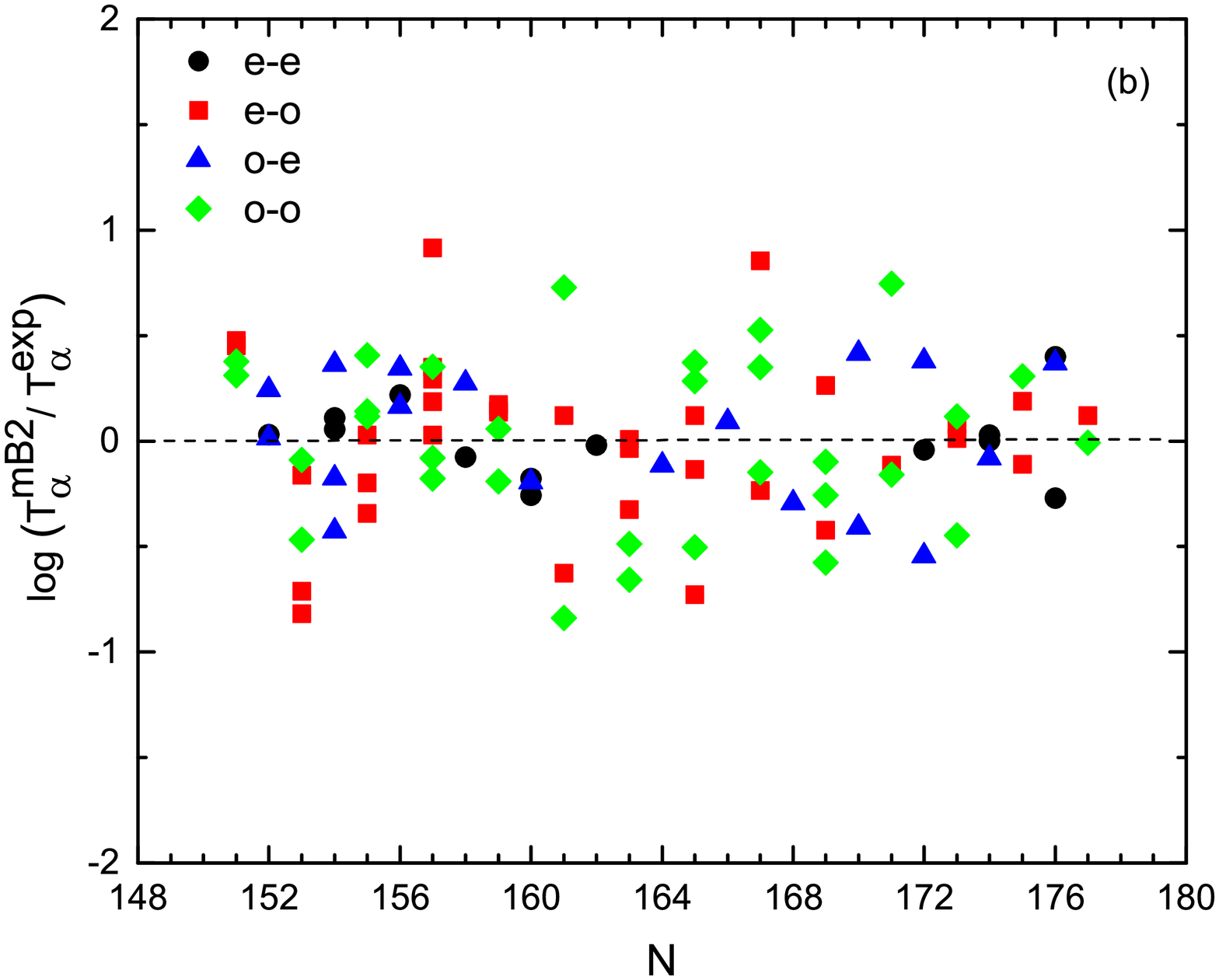}
 \caption{The deviations of the R (a) and mB2 (b) formula predictions from experimental data as function of neutron number $N$ for 96 data points.}
 \label{REXPB2}
\end{figure}

Coming back to the results of the fits, one observes that mB1 fit have almost the same rms values as that corresponding to VS formula fit but which uses in total 6 parameters instead of 7. This is the first proof for the superior predictive power of the proposed modified Brown formula. Moreover, the corresponding rms values of the mB2 fits per parity are better for $(e-e)$ and $(o-e)$, poorer for $(o-o)$ and essentially the same for $(e-o)$ nuclei when compared with the R formula which uses the same number of parameters.

In order to ascertain the overall agreement with experimental data of either two empirical estimations for each fitting scheme, one plotted in Figs. \ref{VSEXPB1} and \ref{REXPB2} the corresponding deviations as function of the neutron number which assures a larger spread of data points. From these plots one can extract some useful information. For example, from Fig. \ref{VSEXPB1} one can see that there are more nuclei which are better described by VS formula. However the nuclei which are better described by mB1 formula have much smaller deviations than those corresponding to VS calculations. Moreover, the deviations of both formulas have the same sign apart from few exceptions. The situation per parity in what concerns the comparison of mB1 and VS results is as follows: the $(o-o)$ nuclei are overall better reproduced by mB1; there is a swap in the best theoretical reproduction for $(e-o)$ and $(o-e)$ nuclei near $N=162$ neutron shell closure, {\it i.e.} the nuclei with $N<162$ are described better by the mB1, and otherwise for the rest of nuclei of the same parity type. As a matter of fact the major differences between Fig. \ref{VSEXPB1} (a) and (b) are recorded mostly in its $N<162$ part. A similar trend is found also in the same plot but for R and mB2 formulas depicted in Fig.\ref{REXPB2}. Here, the most distinct distribution of points is associated to $e-o$ nuclei because of large rms values of both formulas, and in $(o-e)$ nuclei due to the highest difference between associated rms values. The largest discrepancies between the corresponding deviations are associated to few $(o-e)$ and $(o-o)$ nuclei whose empirical description favors the R formula. Leaving aside these singular cases, the plot shows that the deviations from experiment of R and mB2 formulas are more or less identical for the majority of the even $A$ nuclei.

The above analysis points to the fact that there are regions, in not yet explored part of the nuclide chart, where the considered formulas can give almost similar or largely different predictions. The first situation obviously provides values with high probability of experimental realization, while the second case can at least provide an existence interval where the true value is situated.

\subsection{Generating predictions}

\begin{figure}[ht!]
 \centering
 \includegraphics[width=0.68\textwidth]{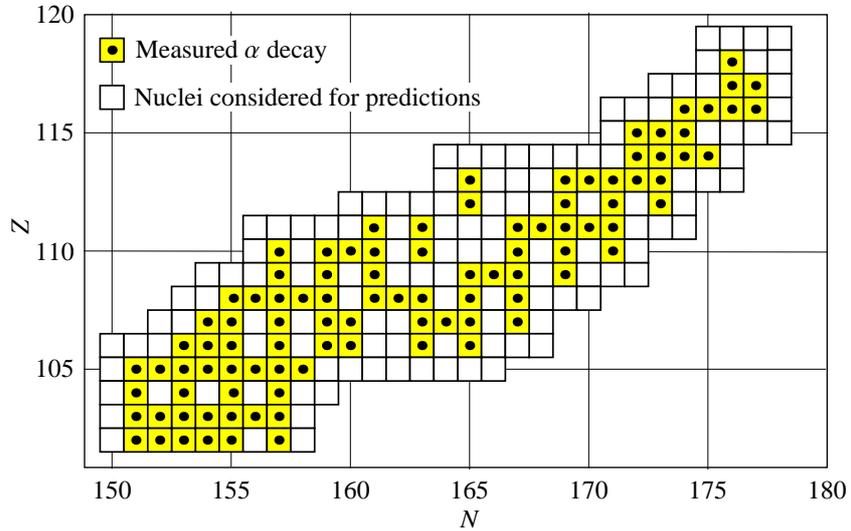}
 \caption{The chart of SHN considered in the study.}
 \label{diag}
\end{figure}

The 125 superheavy nuclei considered for predictions are schematically shown in Fig.\ref{diag}. The chosen region of the nuclide chart ranges between $N=150-178$ and $Z=102-119$. As can be seen, the nuclei chosen for predictions with few exceptions neighbour at least one experimentally measured $\alpha$ emitter. This specific selection of nuclei is related to the $Q_{\alpha}$ formula for predictions. As was shown in Ref. \cite{DongPRL}, the adopted formula not only has an overall good agreement with experimental data, but is also very successful in relating experimental $Q_{\alpha}$ values of neighbouring super heavy nuclei. Thus, in the first stage one generates the predictions for the $Q_{\alpha}$ values by using the five-parameter formula (\ref{MDong}) with parameters taken from \cite{DongPRC,DongPRL}. Predicted $Q_{\alpha}^{Th}$ values are then plugged into the VS, R and the two mB empirical formulas with parameters fixed above and collected in Table \ref{par}, finally obtaining four distinct $\log{T_{\alpha}}$ predictions.

\begin{figure}[ht!]
 \centering
 \includegraphics[clip,trim = 26mm 10mm 35mm 20mm,width=0.68\textwidth]{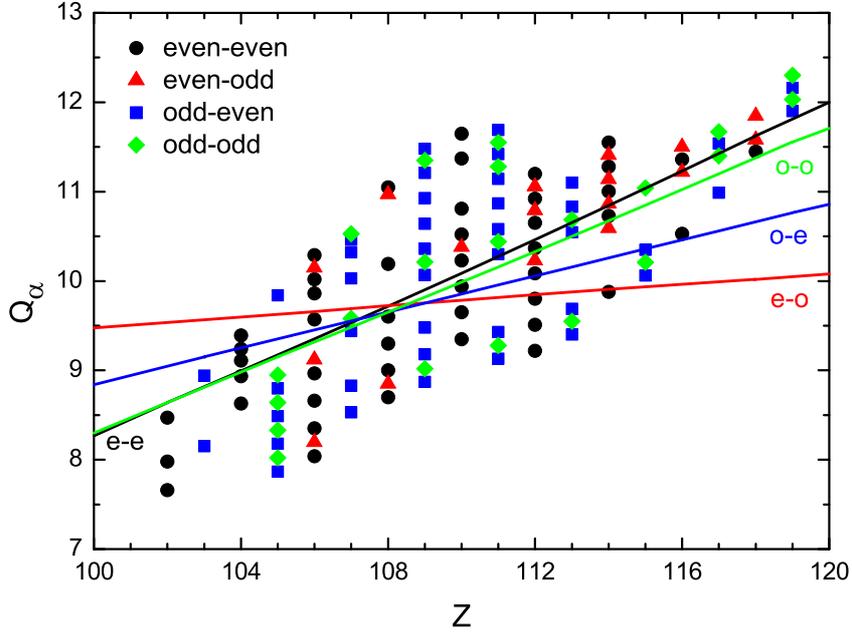}
 \caption{Curves $Q_{\alpha}=f(Z)$ where the $\log{T_{\alpha}}$ values of mB1 and mB2 formulas match are confronted with the predicted 125 $(Z,Q_{\alpha}^{Th})$ data points imparted on the $(Z,N)$ parity.}
 \label{QZ}
\end{figure}

Before sorting the resulted $\log{T_{\alpha}}$ values, it would be opportune to confront analytically the fitting formulas. As the VS, R and mB formulas are distinct in structure we will focus on the two fitting schemes associated to the mB formula which differ only in the involved parameters. Equating the two expression for each parity, one obtains a relation between $Q_{\alpha}$ and $Z$ only. In the plane defined by these quantities, this relation represents a curve following $(Z,Q_{\alpha})$ points for which the two employed fitting expressions provide the same results for $\log{T_{\alpha}}$. These lines are visualized in Fig. \ref{QZ}, where also are distributed the prediction results for the considered 125 nuclei. As can be seen, the lines corresponding to $(e-e)$ and $(o-o)$ parities are almost superposed and cross the most prediction data points $(Z,Q_{\alpha}^{Th})$. This is also consistent with the results of the experimental data fits, where the mB1 and mB2 perform better for these particular parities. Another interesting fact is that all curves share the same intersection point around $Z=108$, which is incidentally a magic number.

\begin{figure}[bt!]
 \centering
 \includegraphics[clip,trim = 26mm 10mm 35mm 20mm,width=0.68\textwidth]{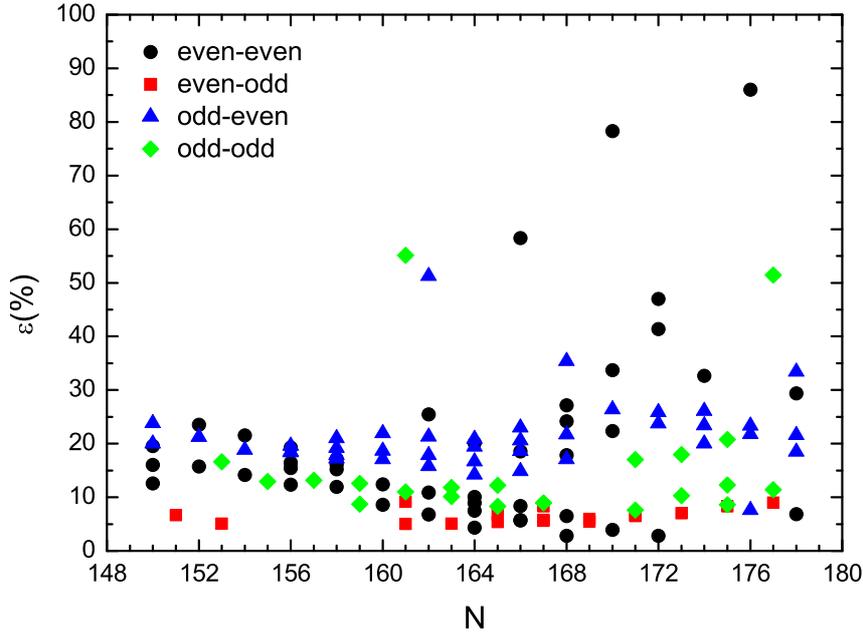}
 \caption{The relative deviation $\varepsilon$ between the predictions of the four fitting formulas is given in percents as function of neutron number $N$ for the selected 125 SHN with unknown $\alpha$ decay properties.}
 \label{eN}
\end{figure}

Judging by the slopes of the curves and the trend of predicted $(Z,Q_{\alpha}^{Th})$ data points one can ascertain that the two mB fitting procedures provides more similar results for higher $Z$ nuclei. However, in order to pinpoint the nuclei with reliable $\log{T_{\alpha}}$ predictions, their analysis must be also corroborated with the results provided by VS and R estimations. To do this, one introduces the following quantity:
\begin{equation}
\varepsilon=\frac{Max\left(\left|\log{T_{\alpha}^{i}}\right|\right)-Min\left(\left|\log{T_{\alpha}^{i}}\right|\right)}{Max\left(\left|\log{T_{\alpha}^{i}}\right|\right)},
\end{equation}
with $i$ indexing the four results corresponding to VS, R, mB1 and mB2. This relative error judges the worthiness of the predictions for each nuclei by relating the spread of the four values with the maximal predicted value. The $\varepsilon$ values for each of the 125 nuclei with unknown $\alpha$ decay properties are plotted as function of neutron number $N$ in Fig. \ref{eN}. This graphical representation facilitates the selection of nuclei with maximally superposing predictions. Although, almost all prediction points have $\varepsilon<40\%$, there are some cases with very high discrepancies, such as for example $^{290}$Fl and $^{280}$Ds nuclei with a relative deviation between different estimations of about 86\% and 78\%. The obvious conclusion which can be drawn from Fig. \ref{eN} is that all $(e-o)$ nuclei have $\varepsilon<10\%$, and therefore the four fitting results converge to approximately the same value. A similarly good agreement between predictions is obtained for selected high $N$ $(e-e)$ and $(o-o)$ nuclei and only for one $(o-e)$ nucleus, $^{291}$115. As a matter of fact, the agreement between various estimations is sensibly improving for $(e-e)$ and $(o-o)$ starting around neutron $N=162$ shell closure. The emergence of turning points at neutron $N=162$ and proton $Z=108$ magic numbers identified above are at first glance just numerical results. However, nuclei with closed shells beside exhibiting an enhanced stability are also critical in what concerns the evolution of nuclear properties with the number of occupied single particle levels. Therefore, the fact that such transitions are registered, even when using the present empirical formulas, supports the procedure of fitting experimental data on a specific region of the nuclide chart, more precisely of the SHN.

The predictions obtained with the four fitting formulas for all these nuclei with $\varepsilon<10\%$ are listed in Table \ref{tabp} together with the associated $\varepsilon$ and average $\log{T_{\alpha}}$ values. In virtue of the present analysis, the averages calculated in Table \ref{tabp} for 36 nuclei represent $\log{T_{\alpha}}$ values with highly probable experimental realisation.

\renewcommand{\arraystretch}{0.68}
\setlength{\tabcolsep}{6.5pt}
\begin{table*}[ph!]
\caption{Theoretically estimated $Q_{\alpha}^{Th}$ values obtained by means of formula (\ref{MDong}) and the corresponding half lives for SHN predicted through the VS, R, mB1 and mB2 empirical formulas with a relative deviation between estimations less than 10\%. For practical reasons the simple arithmetic average of the four values is also presented.}
\label{tabp}
\begin{center}
\begin{tabular}{cccccccccc}
\hline\hline\noalign{\smallskip}
 $Z$& $N$& $A$& $Q_{\alpha}^{Th}$& $\log{T_{\alpha}^{VS}}$& $\log{T_{\alpha}^{mB1}}$& $\log{T_{\alpha}^{R}}$& $\log{T_{\alpha}^{mB2}}$& $\varepsilon$& $\log{T_{\alpha}^{av}}$\\
    &    &    & (MeV)           & (s)                     & (s)                     & (s)                   & (s)                     &(\%)          &  (s)                  \\
\noalign{\smallskip}\hline\noalign{\smallskip}
105 & 159 & 264 & 8.95 & 1.179 & 1.181 & 1.078 & 1.118 & 8.70 & 1.139\\
        \noalign{\medskip}
 106 & 151 & 257 & 10.15 & -1.846 & -1.893 & -1.823 & -1.953 & 6.66 & -1.879\\
 106 & 161 & 267 & 9.12 & 0.641 & 0.622 & 0.666 & 0.685 & 9.20 &  0.654\\
 106 & 167 & 273 & 8.20 & 3.239 & 3.247 & 3.367 & 3.439 & 5.82 & 3.323\\
        \noalign{\medskip}
 108 & 153 & 261 & 10.97 & -3.164 & -3.197 & -3.228 & -3.335 & 5.11 & -3.231\\
 108 & 164 & 272 & 9.60 & -0.558 & -0.613 & -0.597 & -0.562 & 8.97 & -0.583\\
 108 & 169 & 277 & 8.85 & 1.859 & 1.822 & 1.880 & 1.928 & 5.50 & 1.872\\
        \noalign{\medskip}
 110 & 165 & 275 & 10.38 & -1.483 & -1.503 & -1.590 & -1.573 &6.73 & -1.537\\
 110 & 166 & 276 & 10.23 & -1.603 & -1.645 & -1.749 & -1.714 &8.35 & -1.678\\
 110 & 168 & 278 & 9.94 & -0.919 & -0.970 & -0.952 & -0.907 &6.49 & -0.937\\
       \noalign{\medskip}
 111 & 165 & 276 & 10.44 & -1.067 & -1.072 & -0.983 & -1.001 & 8.30 & -1.031\\
       \noalign{\medskip}
 112 & 160 & 272 & 11.20 & -3.271 & -3.257 & -3.490 & -3.563 & 8.59 & -3.395\\
 112 & 161 & 273 & 11.06 & -2.543 & -2.517 & -2.565 & -2.650 & 5.02 & -2.569\\
 112 & 162 & 274 & 10.92 & -2.703 & -2.703 & -2.838 & -2.900 & 6.79 & -2.786\\
 112 & 163 & 275 & 10.79 & -1.961 & -1.949 & -1.980 & -2.054 & 5.11 & -1.986\\
 112 & 164 & 276 & 10.65 & -2.107 & -2.120 & -2.149 & -2.203 & 4.36 & -2.145\\
 112 & 166 & 278 & 10.37 & -1.478 & -1.507 & -1.421 & -1.468 & 5.71 & -1.469\\
 112 & 167 & 279 & 10.23 & -0.704 & -0.721 & -0.709 & -0.768 & 8.33 & -0.726\\
      \noalign{\medskip}
 113 & 167 & 280 & 10.69 & -1.192 & -1.170 & -1.085 & -1.126 & 8.98 & -1.143\\
          \noalign{\medskip}
 114 & 164 & 278 & 11.55 & -3.594 & -3.529 & -3.778 & -3.814 & 7.47 & -3.679\\
 114 & 165 & 279 & 11.41 & -2.876 & -2.801 & -2.923 & -2.961 & 5.40 & -2.890\\
 114 & 166 & 280 & 11.28 & -3.047 & -3.000 & -3.151 & -3.180 & 5.66 & -3.095\\
 114 & 167 & 281 & 11.14 & -2.315 & -2.260 & -2.362 & -2.394 & 5.59 & -2.333\\
 114 & 168 & 282 & 11.00 & -2.472 & -2.445 & -2.491 & -2.515 & 2.78 & -2.481\\
 114 & 169 & 283 & 10.87 & -1.726 & -1.691 & -1.770 & -1.798 & 5.95 & -1.746\\
 114 & 170 & 284 & 10.73 & -1.868 & -1.862 & -1.795 & -1.816 & 3.91 & -1.835\\
 114 & 171 & 285 & 10.59 & -1.107 & -1.093 & -1.146 & -1.171 & 6.66 & -1.129\\
          \noalign{\medskip}
 115 & 171 & 286 & 11.04 & -1.566 & -1.500 & -1.447 & -1.455 & 7.60 & -1.492\\
 115 & 176 & 291 & 10.35 & -0.207 & -0.207 & -0.224 & -0.216 & 7.59 & -0.214\\
           \noalign{\medskip}
 116 & 171 & 287 & 11.50 & -2.664 & -2.554 & -2.731 & -2.715 & 6.48 & -2.666\\
 116 & 172 & 288 & 11.36 & -2.832 & -2.753 & -2.819 & -2.808 & 2.79 & -2.803\\
 116 & 173 & 289 & 11.22 & -2.097 & -2.012 & -2.164 & -2.148 & 7.02 & -2.105\\
          \noalign{\medskip}
 117 & 175 & 292 & 11.40 & -1.934 & -1.812 & -1.798 & -1.768 & 8.58 & -1.828\\
          \noalign{\medskip}
 118 & 175 & 293 & 11.85 & -3.010 & -2.833 & -3.089 & -3.020 & 8.30 & -2.988\\
 118 & 177 & 295 & 11.58 & -2.463 & -2.316 & -2.545 & -2.479 & 8.99 & -2.451\\
 118 & 178 & 296 & 11.45 & -2.628 & -2.515 & -2.495 & -2.448 & 6.85 & -2.522\\
 \noalign{\smallskip}\hline\hline
\end{tabular}
\end{center}
\end{table*}

\section{Conclusions}

One of the main result of the present study is the modification of the Brown formula by considering the power of $Z_{d}$ as a fitting parameter. Although such an approach was used to describe cluster radioactivity \cite{Horoi} it was never reconsidered for the $\alpha$-decay. Here one chosen to fix the parameters by fitting just the region of known SHN, which will enable the final formula to predict more probable $\log{T_{\alpha}}$ values for immediate unknown superheavy $\alpha$ emitters. Fitting experimental data for 96 SHN with two distinct schemes showed a variation of the involved parameters from the original ones. This is a new result in view of the fact that systematic comparisons of empirical correlation are usually made on the original parameters \cite{Wang} values which are often old and have a more global character. A special emphasis is made on the diminished power parameter, which is more closer to the 0.5 value instead of the 0.6 power established by Brown \cite{Brown} and later reconfirmed in Ref. \cite{Horoi}. The two fitting schemes are established in such a way as to be compared with two most successful empirical correlations, which are Viola-Seaborg and Royer formulas. Similar fits with these formulas where performed on the same 96 SHN and the results where dully compared with those corresponding to the two modified versions of the Brown formula, mB1 and mB2. The comparison of the fitting results revealed a similar predictive power of the proposed empirical formulas. Nevertheless, the distribution of the deviations from experimental values have some distinct local features which were pointed out through convenient graphical representations. VS formula fares slightly better than mB1, but uses in total 7 parameters in comparison to the 6 of the mB1 fitting procedure. On the other hand, the parity differentiated fits of mB2 are better than those of R formula for $(e-e)$ and $(o-e)$ nuclei and poorer for the rest of parities. Both schemes totals 12 fitting parameters.

As one of the aims is to provide predictions for $\alpha$-decay properties of SHN, the resulting fitting expressions are extrapolated to 125 unknown $\alpha$-emitters adjacent to experimentally measured ones.
In the first step one calculated $Q_{\alpha}$ values based on a simple five-parameter formula. The obtained results are further used as input for corresponding $\alpha$ half live predictions with the VS, R, mB1 and mB2 formulas, whose parameters where fixed by fitting the existent experimental data on $\alpha$ decay of SHN. Numerical analysis of the relative deviation between the four predictions offer surprising regularities in vicinity of proton and neutron shell closures $Z=108$, respectively $N=162$. Moreover, it also serves for the identification of regions where the four $\log{T_{\alpha}}$ results maximally or minimally diverge. Finally, we selected a number of 36 predictions with exceptional superposition of predicted values for which one tabulated the relevant results, including an average $\log{T_{\alpha}^{av}}$ value. These are predominantly nuclei with $Z=$ even and $N=$ odd.

In conclusion, the obtained predictions for the selected nuclei can be regarded as confident reference points for experimental design given the fact that the numbers are obtained by four independent expressions.

\paragraph{Acknowledgments}
The authors acknowledge the financial support received from the Romanian Ministry of Education and Research, through the Projects PN-16-42-01-01/2016 and PN-II-ID-PCE-2011-3-0068.

\end{document}